\documentclass[conference]{IEEEtran}
\IEEEoverridecommandlockouts
\usepackage{cite}
\usepackage{amsmath,amssymb,amsfonts}
\usepackage{algorithmic}
\usepackage{graphicx}
\usepackage{textcomp}
\usepackage{xcolor}
\usepackage{hyperref}
\usepackage{tcolorbox}

\usepackage{graphicx}
\usepackage{balance}
\usepackage{changepage}
\usepackage{subcaption}
\usepackage{enumitem}
\usepackage{bsymb}  
\usepackage{comment}
\usepackage{amsmath}
\usepackage {tikz}
\usetikzlibrary {positioning}
\usepackage[pdf]{graphviz}
\usepackage{array}
\usepackage{xspace}

\definecolor{ebkw}{RGB}{142,38,4}
\definecolor{ebset}{RGB}{0,128,0}
\definecolor{ebb}{RGB}{35,35,194}
\definecolor{ebtag}{RGB}{110,110,120}

\newcommand{\eb}[1]{\textcolor{ebb}{\textsl{#1}}}
\newcommand{\kw}[1]{\textcolor{ebkw}{\textsf{\bf{#1}}}}
\newcommand{\ebtag}[1]{\textcolor{ebtag}{\textit{#1}}}
\newcommand{\ebcset}[1]{\textcolor{ebset}{#1}}

\newcommand\mhr{\texttt{MyHR}\xspace}

\def\BibTeX{{\rm B\kern-.05em{\sc i\kern-.025em b}\kern-.08em
    T\kern-.1667em\lower.7ex\hbox{E}\kern-.125emX}}
\begin{document}

\title{Formal Verification of Access Control Model for \textit{My Health Record} System}
\author{\IEEEauthorblockN{Victor Rivera}
\IEEEauthorblockA{
\textit{Australian National University}\\
Canberra, Australia \\
victor.rivera@anu.edu.au}
}

\maketitle
\thispagestyle{plain}
\pagestyle{plain}
\begin{abstract}
My Health Record system is the Australian Government's digital health record system that holds \textit{My Health Record}. \textit{My Health Record} is a secure online health record containing consumers' health information. The system aims to provide health care professionals with access to key health information, e.g. listing medicines, allergies and key diagnoses; radiology and pathology test results. The system (previously named   \textit{Personally   Controlled   Electronic Health Record}) enables consumers to decide how to share information with any of their health care providers who are registered and connected to the system. The \textit{My Health Record} system operates under the Australian legislative framework \textit{My Health Records Act 2012}. The Act establishes, inter alia, a privacy framework specifying which entities can collect, use and disclose certain information in the system and the penalties that can be imposed on improper collection, use and disclosure of this information. This paper presents the formal specification (from the legislation) and verification of the \textit{My Health Record} regarding how consumers can \textit{control} who access the information, and how the system adheres to such \textit{access}. We rely on the correct-by-construction Event-B method to prove \textit{control} and \textit{access} properties of the system. 

\end{abstract}


\section{Introduction}
The interest in introducing electronic personal health record systems has emerged as an important area of research in the medical and health informatics domain. These systems are indeed complex to understand, implement and use as they have many different components interacting to each other while managing sensitive information of their users. An electronic personal health record system is a digital version of a patient's health information. The system makes patient's information available instantly and securely to authorised users. Many countries have implemented their own system: the National Health Service implemented the Summary Care Record\cite{scr} in UK, ELGA\cite{web:elga} (\textit{elektronische Gesundheitsakte}) in Austria,  the National Electronic Health Record\cite{nehr} in Singapore, My Health Records\cite{mhr} in Australia, among others. 

In Australia, the first national electronic health record system was launched in 2012\cite{Xu:13}, initially named Personally Controlled Electronic Health Record and later changed (in 2016) to My Health Record. The system is supported by a legislative framework. The objectives of the system were defined in the \textit{My Health Records Act 2012 (Cth)}\footnote{Retrieved from \href{https://www.legislation.gov.au/Details/C2019C00337}{https://www.legislation.gov.au/Details/C2019C00337}} by the Australian government as follows:

`` to enable the establishment and
operation of a voluntary national system for the provision of
access to health information relating to recipients of
healthcare, to:
\begin{enumerate}
    \item Help overcome the fragmentation of health information; and
    \item Improve the availability and quality of health information; and
    \item Reduce the occurrence of adverse medical events and the duplication of treatment; and
    \item Improve the coordination and quality of healthcare provided to healthcare recipients by different healthcare providers.''
\end{enumerate}

The system is different from others implemented in other countries as its emphasis is to empower Australians through being able to personally control their health information: what is displayed and how others (e.g. GPs) can access such information. Consistent with this view, the government chose an opt-in model: users must register to use the system. However, the opt-in model did not achieve the expected number of users. It is reported that only 803 people registered in the first week, 4500 people in the first month, and as low as 6000 in the first two months all over Australia; while the targeted user number is 500000 within the first year of the operation of the system\cite{Xu:13}. In 2019, the government  adopted an opt-out model: all Australians were automatically registered and the decision of non-participating has to be specially requested by the individuals. This model has impacted the number of users of the system. My Health Record system (as in April, 2020) has 22.7M users, over 15.6M of them have data, around 1.95B health documents have been uploaded, and more than 90\% of pharmacies, hospitals and service providers are registered\cite{statistics}.

Despite letting users to personally control their health information, they have no confidence on the system in terms of their privacy and who can actually access the information and how\cite{concept:operations, Patten:19, Lupton:19}. Privacy is addressed by having users choosing who can or cannot access their records. Access measures are addressed by having laws and regulation in place. However, there is not a clear way to ensure this to users. In this paper, we present a formal specification of the My Health Record system based on the legislation that defines it. We focused on the control and access properties of the health information. We also  present the formal verification of such properties and the validation of the system (using animation of the system and code generation). We believe that the use of formal methodologies to ensure the access and control of My Health Record sensitive information will give users the confidence to properly use the system. Furthermore, this work opens a promising avenue for using a more formal approach to check (verify) legislation. 

This paper is structured as follows. Section \ref{sec:preliminaries} introduces the formal methodologies and the mathematical notation used. Section \ref{sec:system} describes the My Health Record system and its requirements within the access and control framework. The formalisation of the system is introduced in Section \ref{sec:model}. The section shows the formal specification and verification of the system in Event-B\cite{Abrial:06}. And the validation step, animation in ProB\cite{prob}, and code generation, with EventB2Java\cite{CodeGen}. Section \ref{sec:relatedWork} compares our approach with similar works. Finally, Section \ref{sec:conclusion} is devoted for conclusions and future work.

\section{Preliminaries}
\label{sec:preliminaries}
\subsection{Event-B}
Event-B\cite{Abrial:06} is a formal method technique to describe and analyse the behaviour of reactive systems. Event-B language is based on predicate logic and set theory. It includes a full-fledged battery of operations over sets and relations for modelling discrete software systems. Event-B models are composed of machines and contexts. Three basic relationships are used to structure an Event-B model, namely, a machine \textbf{sees} a context or can \textbf{refine} another machine, and a context can \textbf{extend} another context. Machines contain the dynamic parts of a model (i.e. variables, invariants, events), and contexts the static part of a model (i.e carrier sets, constants). Events (after keyword \kw{events} in Figure \ref{fig:example}) are composed of
two parts, the guard (the \kw{WHERE} keyword in Figure \ref{fig:example}) and the actions (the \kw{THEN} keyword  in Figure \ref{fig:example}). Actions can only be executed if the guard holds. Each action determines how a machine variable evolves by modelling a variable assignment in Event-B (the \eb{:=} symbol). In Event-B, systems are typically modelled via a sequence of refinements. First, an abstract machine is developed and verified to satisfy whatever correctness and safety properties are desired. Refinement machines are used to add more detail to the abstract machine until the model is sufficiently concrete for hand or automated translation to code. Refinement proof obligations are discharged to ensure that each refinement is a faithful model of the previous machine, so that all machines satisfy the correctness properties of the original.


\begin{figure}[h!]
{\scriptsize
\centering
\[
\begin{array}{l}
\kw{machine}\eb{ abs\_mch}\kw{ sees}\eb{ ctx\_abs}\\
\hspace*{.5cm}\kw{variables}\eb{ my\_health\_report\_DB consumer MyHR}\\
\kw{invariants}\\
\hspace*{1cm}\ebtag{inv1}\eb{ my\_health\_report\_DB $\subseteq$ }\ebcset{MY\_HEALTH\_RECORD}\\
\hspace*{1cm}\ebtag{inv2}\eb{ consumer $\subseteq$ }\ebcset{PEOPLE}\\
\hspace*{1cm}\ebtag{inv3}\eb{ MyHR $\in$ consumer $\tbij$  my\_health\_report\_DB}\\
\kw{events}\\
\hspace*{.5cm}\eb{opt\_out} \triangleq\\
\hspace*{1cm}\kw{ANY}\eb{ c}\\
\hspace*{1cm}\kw{WHERE}\\
\hspace*{1.5cm}\ebtag{grd1}\eb{ c $\in$ consumer}\\
\hspace*{1cm}\kw{THEN}\\
\hspace*{1.5cm}\ebtag{act1}\eb{ consumer := consumer $\backslash$ \{c\}}\\
\hspace*{1.5cm}\ebtag{act2}\eb{ my\_health\_report\_DB := my\_health\_report\_DB $\backslash$ \{MyHR(c)\}}\\
\hspace*{1.5cm}\ebtag{act3}\eb{  MyHR := MyHR $\backslash$ \{c$\mapsto$MyHR(c)\}}\\
\hspace*{1cm}\kw{END}\\
\kw{END}
    \end{array}
  \]
}
\caption{Excerpt of My Health Record system model in Event-B.}
\label{fig:example}
\end{figure}

Figure \ref{fig:example} is an excerpt of the My Health Record system in Event-B. Machine \eb{abs\_mch} sees (keyword \kw{sees}) context \eb{ctx\_abs}. The context is omitted in the figure; it defines two carrier sets (new data type) \ebcset{MY\_HEALTH\_RECORD} modelling all possible record spaces in the system, and \ebcset{PEOPLE} modelling all possible people of the system. Machine \eb{abs\_mch} defines three variables: \eb{my\_health\_report\_DB}, the set of all record spaces currently active in the system; \eb{consumer}, the set of all users of the system; and \eb{MyHR}, mapping consumers to their record space. \eb{MyHR} is defined as a one-to-one function (\eb{$\tbij$}). The possible values of variables is shaped by the invariant (after keyword \kw{invariants}). Event \eb{opt\_out} models the action of a consumer \eb{c} (after keyword \kw{ANY}) to opt out the system. Consumers can opt out at any time, as long as they are part of the system (\ebtag{grd1}). If the event is triggered, the consumer is removed from the set of active users (\ebtag{act1}), as well as the consumer's My Health Record (\ebtag{act2}). To maintain consistency with the invariant (\ebtag{inv3}), the mapping \eb{c$\mapsto$MyHR(c)} is removed from \eb{MyHR} (\ebtag{act3}). In Event-B, actions are executed in parallel.

Rodin\cite{Abrial:10} is an open source Eclipse based integrated
development environment (IDE) for Event-B model development. The Rodin is a core set of plug-ins for project management, formal development, syntactic analysis, proof assistance and proof-based verification. Moreover, it comes with additional plugins to provide different functionalities and features: ProB\cite{prob} translates Event-B models to B to model finding, checking, deadlock and test case generation. It can also be used to animate the model; EventB2Java\cite{CodeGen} is a code generator plugin that soundly generates JML-annotated Java code from Event-B. EB2ALL\cite{Dominique:11} is a plug-in that includes the EB2C, EB2C++, EB2J, and EB2C\# plug-ins, each translating Event-B machines to the indicated language.


\subsection{Mathematical notations}
The following is the mathematical notation underlying our model. It is drawn from the Event-B notation\cite{Abrial:06}.

\subsubsection{Sets}
Let $S$ and $T$ be sets and $E$ and $F$ expressions:
\begin{itemize}
    \item $\emptyset$ denotes the empty set.
    \item $S \bunion T$ is the set union.
    \item $S \binter T$ is the set intersection.
    \item $S \backslash T = \{x \mid x \in S \wedge x \not \in T\}$ is the set difference.
    \item $\pow (S) = \{x \mid x \subseteq S\}$, the power set of $S$.
    \item $E \mapsto F$ is an ordered pair.
    \item $S \times T = \{x \mapsto y \mid x \in S \wedge y \in T\}$ is the Cartesian product.
\end{itemize}

\subsubsection{Set predicates}
Let $S$, $T$ and $U$ be sets:
\begin{itemize}
    \item $S \subseteq T$ denotes subset.
    \item $partition(U, S, T)$, $S$ and $T$ partition the set $U$, i.e.
    $U = S \bunion T \wedge S \binter T = \emptyset$.
\end{itemize}
\subsubsection{Relations}
A relation is a set of ordered pairs; a many to many mapping. Let $S$ and $T$  be sets and $r$ and $p$ relations:
\begin{itemize}
    \item $S \rel T = \pow (S \times T)$ is a relation.
    \item $dom(r) = \{x \mid \exists y \cdot x \mapsto y \in r\}$ denotes the domain of $r$.
    \item $ran(r) = \{y \mid \exists x \cdot x \mapsto y \in r\}$ denotes the range of $r$.
    \item $r;p = \{x \mapsto y \mid \exists z \cdot x \mapsto z \in r \wedge z \mapsto y \in p\}$ denotes forward composition.
    \item $S \domres r = \{x \mapsto y \mid x \mapsto y \in r \wedge x \in S\}$ denotes domain restriction.
    \item $r \ranres T = \{x \mapsto y \mid x \mapsto y \in r \wedge y \in T\}$ denotes range restriction.
    \item $S \ranres id = \{x \mapsto x\mid x \in S\}$ denotes the identity.
    \item $r^{-1} = \{y \mapsto x\mid x \mapsto y \in r\}$ is the inverse of relation $r$.
    \item $r[S] = \{y\mid \exists x \cdot x \in S \wedge x \mapsto y \in r\}$ denotes relation image.
\end{itemize}

\subsubsection{Functions}
A function is a relation with the restriction that each element of the domain is related to a unique element in the range; a many to one mapping. Let $S$ and $T$ be sets:
\begin{itemize}
    \item $S \pfun T = \{r \mid r \in S \rel T \wedge r^{-1};r \subseteq T \ranres id\}$ denotes partial function.
    \item $S \tfun T = \{f \mid f \in S \pfun T \wedge dom(f) = S\}$ denotes total function.
    \item $S \tbij T = \{f\mid f \in S \tfun T \wedge f^{-1} \in T \pfun S \wedge ran(f) = T\}$ denotes a bijective function; a one to one relation.
\end{itemize}

\section{My Health Record System}
\label{sec:system}
My Health Record system (herein \mhr) is an online centralised folder summary documents relating to the healthcare of consumers. The system is similar to a Dropbox folder\footnote{\href{www.dropbox.com}{www.dropbox.com}}
where the health information of a consumer is summarised and can be accessed by different entities. The system aims to: improve consumers' care, safety and medical communication. The lack of interoperability between hospitals and GPs is a common source of medical error. \mhr allows for health information to be in one place, which aims to avoid this problem; reduce the need to recite medical history. This reduces the need for consumers to explain (which could be wrong) their medical history; improve continuity of care between providers. Consumers often visit different service providers for the same condition. Having all health information consolidated in one single place might help GPs to continue a treatment.

\begin{figure}[h]
\centering
\includegraphics[width=3.5in]{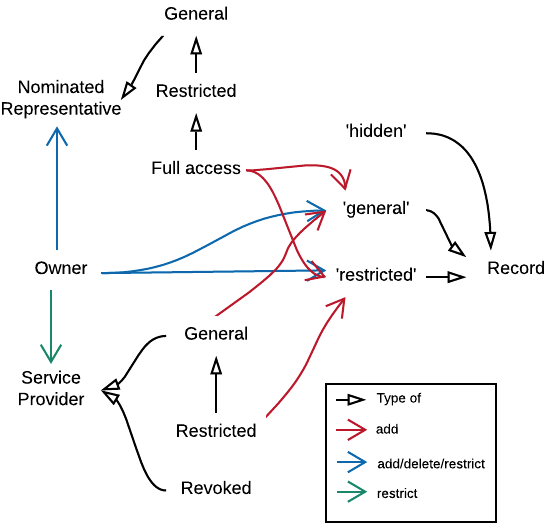}
\caption{Control scheme in \mhr}
\label{fig:control}
\end{figure}

The folder is mainly controlled by consumers, so, in principle, they can decide who can access which information. The system enables consumers to access their health information, add records (e.g. medication the consumer is taking), set privacy controls, hide or remove information and access another consumer's \mhr (see nominated and authorised representatives below). This section introduces the system based on the Australian Acts \textit{My Health Records Act 2012 (Cth)}\footnote{Retrieved from \href{https://www.legislation.gov.au/Details/C2019C00337}{https://www.legislation.gov.au/Details/C2019C00337}} and \textit{My Health Records Rule 2016 (Cth)}\footnote{Retrieved from \href{https://www.legislation.gov.au/Details/F2016C00607}{https://www.legislation.gov.au/Details/F2016C00607}}. The emphasis is on the \textit{control} and \textit{access} properties of the system. Figures \ref{fig:control} and \ref{fig:access} show the schematic representation of control and access of the different elements of the system. Figure \ref{fig:control} shows what can be controlled by the elements of the system. The `Type of' arrow also indicates inheritance of permissions. For instance, a Restricted service provider is a type of General service provider. It can add restricted records as well as general ones. Figure \ref{fig:access} shows the access permissions to information. Similarly, the `Type of' arrow also indicates inheritance of access. For instance, a Full access nominated representative can access both general and restricted records.

\subsection{Definitions}
Some of the definitions used throughout the paper are:
\begin{description}
\item[My Health Record: ] the record of healthcare information that is created and maintained by the \textbf{system operator} in relation to the \textbf{consumer};
\item[My Health Record system (\mhr): ] a system for the collection, use and disclosure of healthcare information, in accordance to the \textbf{consumer}'s wishes;
\item[Record: ] information or an opinion about the health of a \textbf{consumer}. For instance, information about allergies, blood test results, medical conditions, prescriptions. \item[System operator: ] a person in charge of establishing and maintaining the \mhr;
\item[Consumer: ] a person who has received, receives, or may receive healthcare. (Defined as \textit{healthcare recipient} in the Act.);
\item[Service provider: ] a registered entity that provides healthcare. (Defined as \textit{healthcare provider organisation} or \textit{individual healthcare provider} in the Act.);
\item[Nominated Representative: ] a person appointed by the \textbf{consumer} (in agreement with the \textbf{system operator}) to access the \textbf{consumer}'s \mhr, in accordance to the \textbf{consumer}'s wishes;
\item[Authorised Representative: ] a person (in agreement with the \textbf{system operator}) who has parental responsibility for a \textbf{consumer} aged under 14 or for a \textbf{consumer} who is not capable of making decisions for themselves. 
\end{description}

\subsection{The system}
Consumers, once part of the system, are entitled to own only one \mhr; ownership cannot be shared. Initially, the \mhr is empty, no health information has been uploaded.  Over time, the \mhr is populated with consumer's health information (i.e. records),  e.g. after a consumer visits their GPs, pharmacists or hospitals. A records is associated to only one \mhr and has only one owner: the owner of the \mhr where the record sits. Records can be marked as one of the following types:

\begin{itemize}
    \item `general':  records that can be accessed by anyone with access permission to the \mhr. Records are marked `general' by default;
    \item `restricted': restricted records that need a special permission to be accessed;
    \item `hidden': no one (including its owner) has access to these records. Consumers can recover them in agreement with the system operator.
\end{itemize}

Consumers can access their own records marked as `general' and `restricted'. A key feature of the system is that consumers can add and delete records, and control which service providers and nominated representatives access what records. 

\begin{figure}[h]
\centering
\includegraphics[width=3.5in]{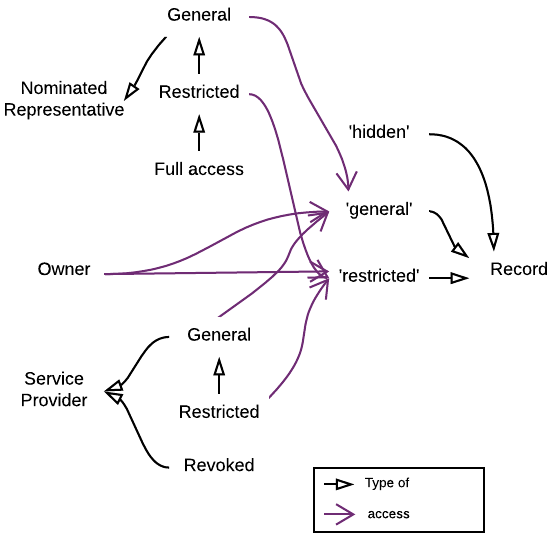}
\caption{Access scheme to records in \mhr}
\label{fig:access}
\end{figure}

Service providers can be in charge of consumers' healthcare. This gives them access to consumers' \mhr. As part of the control given to consumers, they can list their service providers as \textit{General}, \textit{Restricted} or \textit{Revoked}. Thus, consumers restrict the access of service providers to specific records: \textit{General} service providers can access any record marked as `general'. They can also upload records to the \mhr. Records will be marked as `general'; \textit{Restricted} service providers can access any record marked as `general' or `restricted'. (\textit{General} access permissions are included in the \textit{Restricted} access permissions.) In addition, \textit{Restricted} service providers can upload records to the \mhr. Records will be marked as `restricted'; \textit{Revoked} service providers have access to no records.

Consumers can also be nominated or authorised representatives of someone else's \mhr. A nominated representative can access or help manage consumers' \mhr. They might be a family member, close friend or carer. Consumers can add or delete nominated representatives to their \mhr; they cannot nominate themselves. Similar to service providers, consumers can restrict the access of records to nominated representatives by assigning them to specific lists. A nominated representative can be in one of the following lists: \textit{General}, \textit{Restricted} or \textit{Full access}. \textit{General} nominated representatives can access all records of the \mhr marked as `general'; \textit{Restricted} nominated representatives can access all records of the \mhr marked as `general' and `restricted'. \textit{General} access permissions are included in the \textit{Restricted} access permissions; \textit{Full access} nominated representatives are special \textit{Restricted} representatives (thus, they have the same access permissions) with the additional capacity of uploading records to the \mhr. Records will be marked as either `general' or `restricted'.

Consumers can also set an access code to their \mhr on their restricted records. If they want a service provider or nominated representative with \textit{General} access to access restricted records, the consumer can provide the code\footnote{The model in section \ref{sec:model} does not consider this option as providing a \textit{General} service provider or nominated representative with a code is semantically equivalent as granting them  \textit{Restricted} access}.

An authorised representative is a person who manages the \mhr of someone who cannot manage their own. Hence, consumers cannot be authorised representatives of their own \mhr. An authorised representative might manage the record on behalf of a child, or an adult who lacks capacity. They are empowered under Australian law to do anything that the consumer would be able to do. The system operator is in charge of delegating authorised representatives to a consumer. Authorised representatives have full access and control over a \mhr as if they were the owners. Once a \mhr has at least one authorised representative, the owner of such \mhr loses all control over it. Table \ref{table:properties} summarises the properties of the system concerning access and control of the \mhr.

\begin{table*}[h]
\centering
\begin{tabular}{l|p{13cm}}
\textbf{Property ID} & \textbf{Description} \\ \hline \hline
\textit{p1 (control)} & Consumers can add, delete and restrict their own records. \\ \hline
\textit{p2 (access)} & Consumers can access their own records marked as `general' and `restricted'. \\ \hline
\textit{p3 (access)} & Service providers have access to their consumers' \mhr. \\ \hline
\textit{p4 (control)} & Consumers can restrict access of service providers by listing them as \textit{General}, \textit{Restricted} or \textit{Revoked}. \\ \hline
\textit{p5 (access)} & \textit{General} service providers with access to a \mhr can access all of the \mhr's records marked as `general'. \\ \hline
\textit{p6 (control)} & \textit{General} service providers can also upload records to the \mhr. Records will be marked as `general'. \\ \hline
\textit{p7 (access)} & \textit{Restricted} service providers with access to a \mhr can access all of the \mhr's records marked as `general' and `restricted'. \\ \hline 
\textit{p7.i (access)} & \textit{Restricted} access permissions is a subset of General access permissions. \\ \hline 
\textit{p8 (control)} & \textit{Restricted} service providers can also upload records to the \mhr. Records will be marked as `restricted'. \\ \hline
\textit{p9 (access)} & \textit{Revoked} service providers with access to a \mhr cannot access any record. \\ \hline
\textit{p10 (control)} & Consumers can add or delete nominated representatives to their \mhr. \\ \hline
\textit{p11 (control)} & Consumers can restrict access of nominated representatives by listing them as i.e. \textit{General}, \textit{Restricted} or \textit{Full access}. \\ \hline
\textit{p12 (access)} & \textit{General} nominated representatives can access all records of the \mhr marked as `general'. \\ \hline
\textit{p13 (access)} & \textit{Restricted} nominated representatives can access all records of the \mhr marked as `general' and `restricted'. \\ \hline
\textit{p13.i (access)} &  \textit{Restricted} access permissions is a subset of General access permissions. \\ \hline
\textit{p14 (access)} & \textit{Full access} nominated representatives can access all records of the \mhr marked as `general' and `restricted'. \\ \hline
\textit{p15 (control)} & \textit{Full access} nominated representative can also upload records to the \mhr. Records will be marked as either `general' or `restricted'. \\ \hline
\textit{p16 (control)} & Authorised representatives have full access and control over a \mhr as if they were the owner. Once a \mhr has at least one authorised representative, the owner of such \mhr loses all control over it. \\
\end{tabular}
\caption{Control access properties of \mhr}
\label{table:properties}
\end{table*}

Access to the information can be bypassed in emergency situations. The \textit{My Health Records Act 2012 (Cth)} defines where it may be permissible for entities to bypass the access controls. The Act defines an emergency access function, commonly referred as `break glass' function. Such a function can be used when there is a serious threat to the individual's life, health or safety and their consent cannot be obtained; or
there are reasonable grounds to believe that access to the \mhr of that person is necessary to lessen or prevent a serious threat to public health or safety\footnote{`break glass' functionality is not modelled.}.

\section{Formal model in Event-B}
\label{sec:model}
My Health Record system is composed of different elements that interact to each other. The model in Event-B introduces each element gradually in different machines (refinements). The abstract model contains a series of records marked as `general', `restricted' or `hidden'. The abstract model also defines consumers and the access to records, as well as the mechanisms for consumers to control the access. Finally, this machine introduces the system operators. The first refinement introduces service providers and their access to records. It also provides mechanisms for consumers to restrict service providers' access. The second refinement introduces nominated representatives and the access mechanisms of nominated representatives to records. The final refinement introduces authorised representatives and how they control \mhr. Once a \mhr has an authorised representative, they act as owners of the system and the actual owner loses all control over the \mhr. Event-B models typically start with a very abstract machine and each refinement adds just as much information about the system. Due to space, the model present in this paper shows a more compact version of the original model (fewer refinements, but the same information). Both Event-B models (all POs discharged) can be found in \cite{mhr:model}. Table \ref{tab:refinement} shows a description of each refinement. A more detailed description can be found in the coming subsections.

\begin{table}[h!]
  \centering
  \begin{tabular}{p{1.8cm}|p{6cm}}
    \textbf{Machine} & \textbf{Summary} \\ \hline \hline
    Abstract (section \ref{sec:abs})& Health record, consumers, records and their categories: `general', `restricted' and `hidden'. \\ \hline
    Refinement 1  (section \ref{sec:ref1})& Service Providers, their categories and their level of access to \mhr and records\\ \hline
    Refinement 2  (section \ref{sec:ref2})& Nominated Representatives, their categories and their level of access to \mhr and records\\ \hline
    Refinement 3  (section \ref{sec:ref3})&
    Authorised Representative and their level of access to \mhr and records
    \end{tabular}
  \caption{Refinement strategy summary}
  \label{tab:refinement}
\end{table}


\subsection{My Health Record system (abstract machine)}
\label{sec:abs}
The abstract model defines the fundamental elements of the system: 
 \begin{itemize}
     \item consumers. Healthcare recipients of the system; 
\item records. Consumers' medical information. They can be categorised as `general', `restricted' or `hidden'; and 
\item My Health Record online space. It is owned by a consumer and it contains records. 
 \end{itemize}

The context of the model defines a new data type: \ebcset{PEOPLE}, modelling potential users of the system, as well as potential system operators. Data type \ebcset{MY\_HEALTH\_RECORD} models all possible My Health Record of the system. And data type \ebcset{RESOURCES} models the possible records of the system. The machine defines variable \eb{my\_health\_record\_DB $\subseteq$ }\ebcset{MY\_HEALTH\_RECORD}, the set of all \mhr currently active. Variable \eb{consumer $\subseteq$ }\ebcset{PEOPLE} defines the set of all users of the system. Variable \eb{MyHR $\in$ consumer $\tbij$  my\_health\_record\_DB} maps consumers to their \mhr. It is defined as a bijective function as by definition a consumer owns exactly one \mhr, and each \mhr is owned by exactly one consumer. Variable \eb{system\_operator} defines the set of people (no consumers) that serve as system operators: \eb{system\_operator $\subseteq$ }\ebcset{PEOPLE}\eb{$\backslash$consumer}. Records are the resources of the system  containing medical information about a user; they are the elements to be shared. A record could be a blood test, hospital discharge summary, etc. (Our model does not distinguish between different records.). The model defines variable \eb{records $\subseteq$ }\ebcset{RESOURCES} to represent the set of records in the system. 
Variables \eb{records\_mhr $\in$ records $\tfun$ my\_health\_record\_DB} and \eb{consumer\_own\_records $\in$ records $\tfun$ consumer} are defined as total functions to guarantee that every record in the system is in exactly one \mhr, and has only one owner.

Variables \eb{general\_records $\subseteq$ records}, \eb{restricted\_records $\subseteq$ records}, \eb{hidden\_records $\subseteq$ records} are the set of records marked as `general', `restricted' and `hidden', respectively. A record can be in only one of those categories. This fact is modelled as \eb{partition(records, general\_records, restricted\_records, hidden\_records)}.  Consumers own all records on their \mhr,
\begin{align*} 
\forall r \mapsto c \in &consumer\_own\_records \implies\\ 
\forall mhr \in & records\_mhr (r) \implies \\
& c \mapsto mhr \in MyHR.
\end{align*}

We modelled the property via the following invariant
$$\eb{consumer\_own\_records$^{-1}$;records\_mhr $=$ MyHR}.$$ 

The model also defines events for the \textit{control} of the elements in the \mhr (\textit{p1 (control)}). Events \eb{restrict\_record}, \eb{general\_record} and \eb{hide\_record} give owners of the \mhr the opportunity to control the category of each record. Event \eb{unhide\_record} authorises a system operator to restore records which have been previously removed. Events \eb{delete\_record} and \eb{upload\_record} allow owners of the \mhr to delete and add records (respectively). Event \eb{view\_record} defines the access of a consumer \eb{c} to a record \eb{r}. As consumers can access their own records marked as `general’ and `restricted’ (\textit{p2 (access)}), the event can be triggered only if the \eb{c} owns \eb{r} (\eb{r$\mapsto$c $\in$ records\_ownership}) and \eb{r} is not marked as `hidden' (\eb{r $\not \in$ hidden\_records}). 

Figure \ref{fig:restrict} depicts the event for marking a record as `restricted' in the abstract machine. \eb{restrict\_record} enables a consumer \eb{c} to restrict record \eb{r}. The guard of the event establishes that \eb{c} must have ownership over the record (\ebtag{grd1}). If the event is triggered, record \eb{r} is added to the list of records marked as `restricted' (\ebtag{act1}), and taken out from the list of general records (\ebtag{act2}).

\begin{figure}[h!]
\centering
{\scriptsize
\[
    \begin{array}{l}
    \eb{restrict\_record}   \triangleq
    \kw{ANY}\eb{ r c}\\
    \kw{WHERE}\\
    \hspace*{.5cm}\ebtag{grd1: }\eb{r$\mapsto$c $\in$ records\_ownership}\\
    \kw{THEN}\\
    \hspace*{.5cm}\ebtag{act1: }\eb{restricted\_records :=  restricted\_records $\bunion$ \{r\}}\\
    \hspace*{.5cm}\ebtag{act2: }\eb{general\_records :=  general\_records $\backslash$ \{r\}}\\
    \kw{END}
    \end{array}
  \]
}
\caption{Event for marking a record as `restricted'.}
\label{fig:restrict}
\end{figure}

\subsection{Service Providers (refinement 1)}
\label{sec:ref1}
This model (which is a refinement to the model described in section \ref{sec:abs}) introduces service providers and their access to records in \mhr. The model defines a new data type \ebcset{SERVICE\_PROVIDERS} modelling potential registered service providers. Variable \eb{service\_providers $\subseteq$  }\ebcset{SERVICE\_PROVIDERS} defines the set of registered service providers in the system. Variable \eb{consumer\_sp $\in$ consumer $\rel$ service\_providers} maps consumers to service providers. It is defined as a relation to ensure that a service provider can be in charge of several consumers' healthcare and a consumer can be associated with several service providers. Once a service provider is managing a consumer's health care, such service provider has access (limited by the consumer's will) to the consumer's \mhr (\textit{p3 (access)}). This is captured by variable \eb{sp\_MyHR\_access $\in$ service\_providers $\rel$ my\_health\_record\_DB} that maps services providers to My Health Records, and predicate
\begin{align*} 
\forall c \mapsto sp &\in consumer\_sp \implies \\
& sp \mapsto MyHR(c) \in access\_sp\_mhr,
\end{align*}

that is expressed in the model as the following invariant
$$\eb{access\_sp\_mhr = consumer\_sp$^{-1}$;MyHR}.$$

Variables \eb{general\_sp\_list}, \eb{restricted\_sp\_list} and \eb{revoked\_sp\_list} in the model map service providers to \mhr (\eb{service\_providers $\rel$  my\_health\_record\_DB}). These variables modelled the \textit{General}, \textit{Restricted} and \textit{Revoked} lists, respectively. By default, service providers are in the \textit{General} list. A service provider can be in only one of the lists, this is modelled as \eb{partition(sp\_MyHR\_access, general\_sp\_list, restricted\_sp\_list, revoked\_sp\_list)}. 

Variables \eb{general\_sp\_access $\in$ service\_providers $\rel$ records} and \eb{restricted\_sp\_access $\in$ service\_providers $\rel$ records} model the level of access of service providers to records in \mhr. Service providers in \textit{General} lists can access only records marked as `general' (\textit{p5 (access)}),
\begin{align*} 
\forall sp \mapsto r &\in general\_sp\_access \implies r \in general\_records \wedge \\
& sp \mapsto records\_mhr(r) \in general\_sp\_list. 
\end{align*}

We express the property as the following invariant (notice that the access is not defined for service providers in the \textit{Revoked} list (\textit{p9 (access)}))
\begin{align*} 
&\eb{general\_sp\_access =}\\
&\eb{general\_sp\_list;(general\_records $\domres$ records\_mhr)$^{-1}$}.
\end{align*}

Service providers in \textit{Restricted} lists can access records marked as both `general' and `restricted' (\textit{p7 (access)}), 
\begin{align*} 
\forall sp \mapsto r &\in restricted\_sp\_access \implies \\
& r \in general\_records\bunion restricted\_records \wedge \\
&sp \mapsto records\_mhr(r) \in general\_sp\_list\\
&~~~~~~~~~~~~~~~~~~~~~~~~~~~\bunion restricted\_sp\_list.
\end{align*}

We express the property as the following invariant (notice that the access is not defined for service providers in the \textit{Revoked} list (\textit{p9 (access)}))
\begin{align*} 
&\eb{restricted\_sp\_access =(restricted\_sp\_list;}\\
&~~\eb{((general\_records$\bunion$ restricted\_records)$\domres$records\_mhr)$^{-1}$)}\\ &~~~~~~~~~~~~~~~~~~~~~\eb{$\bunion$}\\
&~~\eb{(general\_sp\_list;(general\_records$\domres$ records\_mhr)$^{-1}$)}
\end{align*}

To ensure that service providers with restricted permissions have also general permissions (\textit{p7.i (access)}), 
\begin{align*} 
\forall sp \mapsto mhr &\in general\_sp\_list \implies \\
& sp \mapsto mhr \in restricted\_sp\_list,
\end{align*}

we added the following invariant
$$\eb{general\_sp\_access $\subseteq$ restricted\_sp\_access}.$$

The model extends the events from the refined machine for the control of the elements in the \mhr to update the new variables. 
New control events are introduced to enable consumers to assign service providers to \textit{General}, \textit{Restricted} and \textit{Revoked} lists (\textit{p4 (control)}).  \eb{revoke\_access\_sp} in Figure \ref{fig:revokesp} shows the event for a consumer \eb{c} to revoke the access of service provider \eb{sp} to \mhr \eb{mhr}. The guards of the event establish that \eb{c} owns \mhr \eb{mhr} (\ebtag{grd1\_r1}), \eb{sp} is in charge of \eb{c}'s healthcare (\ebtag{grd2\_r1}), and \eb{sp} is not already in the \textit{Revoked} list for that \mhr(\ebtag{grd3\_r1}). If the event is triggered, the service provider \eb{sp} is added to the \textit{Revoked} list (\ebtag{act1\_r1}) and removed from the \textit{General} or \textit{Restricted} list (\ebtag{act2\_r1} and \ebtag{act3\_r1}). Access permissions for \eb{sp} to \eb{mhr} are also updated: \eb{sp} cannot access any records in \eb{mhr} (\ebtag{act4\_r1} and \ebtag{act5\_r1}).

\begin{figure}[h!]
\centering
{\scriptsize
\[
    \begin{array}{l}
    \eb{revoke\_access\_sp}   \triangleq   	
    \kw{ANY}\eb{ sp mhr c}\\
    \kw{WHERE}\\
    \hspace*{.5cm}\ebtag{grd1\_r1: }\eb{c $\mapsto$ mhr $\in$ MyHR}\\
    \hspace*{.5cm}\ebtag{grd2\_r1: }\eb{c $\mapsto$ sp $\in$ consumer\_sp}\\
    \hspace*{.5cm}\ebtag{grd3\_r1: }\eb{sp$\mapsto$MyHR(c) $\not \in$ revoked\_sp\_list}\\
    \kw{THEN}\\
    \hspace*{.5cm}\ebtag{act1\_r1}: \eb{revoked\_sp\_list := revoked\_sp\_list $\bunion$ \{sp$\mapsto$ mhr\}}\\
    \hspace*{.5cm}\ebtag{act2\_r1}: \eb{restricted\_sp\_list := restricted\_sp\_list $\backslash$ \{sp$\mapsto$mhr\}}\\
    \hspace*{.5cm}\ebtag{act3\_r1}: \eb{general\_sp\_list := general\_sp\_list $\backslash$ \{sp$\mapsto$mhr\}}\\
    \hspace*{.5cm}\ebtag{act4\_r1}: \eb{general\_sp\_access := general\_sp\_access} \\
    \hspace*{1.5cm}\eb{$\backslash$ (
      								(\{sp\} $\domres$ general\_sp\_access) $\ranres$ records\_mhr$^{-1}$[\{mhr\}])}\\
    \hspace*{.5cm}\ebtag{act5\_r1}: \eb{restricted\_sp\_access := restricted\_sp\_access}\\
    \hspace*{1.5cm}\eb{$\backslash$ (
  	 								(\{sp\} $\domres$ restricted\_sp\_access) $\ranres$ records\_mhr$^{-1}$[\{mhr\}])}\\
    \kw{END}
    \end{array}
  \] 
  }
\caption{Event for revoking access to a service provider.}
\label{fig:revokesp}
\end{figure}

The model also introduces new events for the access to records. Event \eb{view\_record\_service\_provider} enables a service provider to access a record. The event makes sure that the corresponding properties are not violating (e.g. a service provider in the \textit{General} list can only access records marked as `general'). Events \eb{upload\_general\_record\_SP} and \eb{upload\_restricted\_record\_SP} enable service provider to upload general and restricted records, respectively, if they are allowed to do so (\textit{p6 (control)} and \textit{(p8 (control))}).

\subsection{Nominated Representatives (refinement 2)}
\label{sec:ref2}
This model (which is a refinement of the model in section \ref{sec:ref1}) introduces nominated representatives and their access permissions on records. 
The model defines variables \eb{general\_nominated $\in$ consumer $\rel$ my\_health\_record\_DB}, \eb{restricted\_nominated  $\in$ consumer $\rel$ my\_health\_record\_DB} and \eb{full\_access\_nominated $\in$ consumer $\rel$ my\_health\_record\_DB} to represent the lists of \textit{General}, \textit{Restricted} and \textit{Full access} nominated representatives. Invariant \eb{general\_nominated $\binter$ restricted\_nominated = $\emptyset$} ensures that a nominated representative is in either list (\textit{Full Access} nominated are special \textit{Restricted} nominated). Consumers cannot be nominated representatives of their own \mhr,
\begin{align*} 
\forall nr \mapsto mhr &\in general\_nominated \\
&~~~~\bunion restricted\_nominated \implies \\
&~~~~~~~~ nr \mapsto mhr \not \in MyHR.
\end{align*}

We express the property as 
$$\eb{(general\_nominated $\bunion$ restricted\_nominated) $\binter$ MyHR = $\emptyset$}.$$

Variables \eb{general\_nominated\_access $\in$ consumer $\rel$ records} and \eb{restricted\_nominated\_access $\in$ consumer $\rel$ records} model the level of access of nominated representatives to records in \mhr. \textit{General} nominated representative can access only records marked as `general' (\textit{p12 (access)}),
\begin{align*} 
\forall nr \mapsto r &\in general\_nominated\_access \implies \\
& r \in general\_records \wedge \\
&nr \mapsto records\_mhr(r) \in general\_nominated.
\end{align*}

We express the property as 
\begin{align*} 
&\eb{general\_nominated\_access =}\\
&\eb{general\_nominated;(general\_records $\domres$ records\_mhr)$^{-1}$}.
\end{align*}

Whereas \textit{Restricted} nominated representatives can access records marked as both `general' and `restricted' (\textit{p13 (access)}),
\begin{align*} 
\forall sp \mapsto r &\in restricted\_sp\_access \implies \\
& r \in general\_records\bunion restricted\_records \wedge \\
&sp \mapsto records\_mhr(r) \in general\_sp\_list\\
&~~~~~~~~~~~~~~~~~~~~~~~~~~~~~~\bunion restricted\_sp\_list.
\end{align*}

We express the property as 
\begin{align*} 
&\eb{restricted\_nominated\_access =(restricted\_nominated;}\\
&~~\eb{((general\_records$\bunion$ restricted\_records)$\domres$records\_mhr)$^{-1}$)}\\ &~~~~~~~~~~~~~~~~~~~~~\eb{$\bunion$}\\
&~~\eb{(general\_nominated;(general\_records$\domres$ records\_mhr)$^{-1}$)}
\end{align*}

\textit{Full access} representatives have the same access permissions as \textit{Restricted} representatives (\textit{p14 (access)}), 

\begin{align*} 
\forall nr \mapsto mhr &\in full\_access\_nominated \implies \\
& nr \mapsto mhr \in restricted\_nominated.
\end{align*}

We express the property as 
$$\eb{full\_access\_nominated $\subseteq$ restricted\_nominated\_access}.$$

To ensure that nominated representatives with \textit{Restricted} level access have also \textit{General} level access (\textit{p13.i (access)}),
\begin{align*} 
\forall nr \mapsto r &\in general\_nominated\_access \implies \\
& nr \mapsto r \in restricted\_nominated\_access,
\end{align*}

we added the following property 
$$\eb{general\_nominated\_access $\subseteq$ restricted\_nominated\_access}.$$

The model extends the events from the refined machine for the control of the elements in the \mhr to update the new variables. New events are introduced to enable users (owners) to add and delete nominated representatives (\textit{p10 (control)}), as well as to control their level of access. Consumers are enabled to assign nominated representatives  to \textit{General}, \textit{Restricted} or \textit{Full access} (\textit{p11 (control)}). \textit{Full access} nominated can also upload records to \mhr. The model introduces events \eb{upload\_general\_record\_nominated} and \eb{upload\_restricted\_record\_nominated} that enable \textit{Full access} representatives to upload records marked as `general' or `restricted' (\textit{p15 (control)}). 

Event \eb{grant\_full\_access\_to\_nominated} in Figure \ref{fig:grant:fullaccess} shows the event for a consumer \eb{c} to grant \textit{Full access} permissions to nominated representative \eb{n} in \mhr \eb{mhr}. The guards of the event establish that the consumer \eb{c} is not nominating themselves as representative nominated (\ebtag{grd1\_r2}), as the owner of the \mhr cannot be a representative of it; \eb{c} owns \mhr \eb{mhr} (\ebtag{grd2\_r2}); and that \eb{n} is not already a representative nominated of \mhr (\ebtag{grd3\_r2}, \ebtag{grd4\_r2} and \ebtag{grd5\_r2}). If the event is triggered, the nominated representative is added to both lists \eb{restricted\_nominated} and \eb{full\_access\_nominated} (\ebtag{act1\_r2} and \ebtag{act2\_r2}). As \textit{Full access} nominated representatives are a special type of \textit{Restricted} ones, access permissions of \eb{n} in \eb{mhr} are also updated (\ebtag{act3\_r2}): \eb{n} has access to all records in \eb{mhr} that are marked as `general' and `restricted'.

\begin{figure}[h!]
\centering
{\scriptsize
\[
    \begin{array}{l}
    \eb{grant\_full\_access\_to\_nominated}   \triangleq   	
    \kw{ANY}\eb{ n mhr c}\\
    \kw{WHERE}\\
    \hspace*{.5cm}\ebtag{grd1\_r2: }\eb{c $\not =$ n}\\
  	\hspace*{.5cm}\ebtag{grd2\_r2: }\eb{c$\mapsto$mhr $\in$ MyHR}\\
  	\hspace*{.5cm}\ebtag{grd3\_r2: }\eb{n$\mapsto$mhr $\not \in$ general\_nominated}\\
  	\hspace*{.5cm}\ebtag{grd4\_r2: }\eb{n$\mapsto$mhr $\not \in$ restricted\_nominated}\\
  	\hspace*{.5cm}\ebtag{grd5\_r2: }\eb{n$\mapsto$mhr $\not \in$ full\_access\_nominated}\\
    \kw{THEN}\\
    \hspace*{.5cm}\ebtag{act1\_r2: }\eb{restricted\_nominated :=  restricted\_nominated $\bunion$ \{n$\mapsto$mhr\}}\\
  	\hspace*{.5cm}\ebtag{act2\_r2: }\eb{full\_access\_nominated := full\_access\_nominated $\bunion$ \{n$\mapsto$mhr\}}\\
  	\hspace*{.5cm}\ebtag{act3\_r2: }\eb{restricted\_nominated\_access := restricted\_nominated\_access $\bunion$}\\
  	\hspace*{.5cm}\eb{(\{n\}$\times$dom((general\_records$\bunion$restricted\_records)$\domres$(records\_mhr$\ranres$\{mhr\})))}\\
    \kw{END}
    \end{array}
  \] 
  }
\caption{Event for granting full access to a nominated.}
\label{fig:grant:fullaccess}
\end{figure}

The model also introduces new events for the access to records. \eb{view\_record\_nominated} enables a nominated representative to access a record. The event makes sure that the corresponding properties are not violating (e.g. a nominated representation in the \textit{Restricted} list can access records marked as `general' and `restricted'). 

\subsection{Authorised Representatives (refinement 3)}
\label{sec:ref3}
This model (which is a refinement of the model in section \ref{sec:ref2}) introduces authorised representatives and their access and control permissions on records and \mhr. Authorised representatives act as owners of the \mhr. The model defines variable \eb{authorised\_rep $\in$ consumer $\rel$ my\_health\_record\_DB} that maps consumers with authorised privileges to \mhr. Consumers cannot be authorised representatives of their own \mhr,
\begin{align*} 
\forall c \mapsto mhr &\in authorised\_rep \implies 
 c \mapsto mhr \not \in MyHR.
\end{align*}

We express the property as 
$$\eb{authorised\_rep $\binter$ MyHR = $\emptyset$}.$$

If a \mhr has an authorised representative, the owner loses all control of the system, and the authorised representatives take over. The model captures this by adding a guard to each event that an owner can perform ensuring that such \mhr does not have any authorised representative (\textit{p16 (control)}). As an example, guards for events \eb{restrict\_record} (in Figure \ref{fig:restrict}), \eb{revoke\_access\_sp} (in Figure \ref{fig:revokesp}), and \eb{grant\_full\_access\_to\_nominated} (in Figure \ref{fig:grant:fullaccess}) are extended to hold also the following guard \eb{MyHR(c) $\not \in$ ran(authorised\_rep)}: the owner of the \mhr can perform these actions only if the \mhr does not have an authorised representative.

\subsection{Correctness of the model}
We have proved the soundness of the model by discharging all proof obligations (POs) generated by Rodin. The model is composed of an abstract machine and 3 refinements. Our modelling resulted in 525 proof obligations, where about 77\% (406) were discharged automatically using Rodin's built-in provers. This high percentage of automation depends on the modelling style applied. For example, we added some redundant information (as variables in the model) to help provers in the verification process. For instance, variable  \eb{sp\_MyHR\_access}, in the first refinement, maps service providers to \mhr that they have access to. However, this information can be inferred from variables mapping service providers to records (\eb{general\_sp\_access} and \eb{restricted\_sp\_access}) -- as records, by definition, belong to only one \mhr. The decision of adding this redundant information is that the model is self-explained and proof obligations are easier to discharge.

We have proved all properties described in Table \ref{table:properties} (and depicted in figures \ref{fig:control} and \ref{fig:access}). That is to say, we have proved that the model lets: 

\begin{itemize}
    \item consumers have access and control over records (they can decide to mark them as `general', `restricted' or `hidden'); 
    \item consumers have control over service providers. Consumers can list them as \textit{General}, \textit{Restricted} or \textit{Revoked}; 
    \item consumers have also control over nominated representatives. Consumers can add and delete, and list them as \textit{General}, \textit{Restricted} or \textit{Full access}; 
    \item authorised representatives to take full control of the \mhr;
    \item nominate representatives with \textit{Full access} to upload records marked as `general' or `restricted';
    \item nominated representatives with \textit{Full access} and \textit{Restricted} permissions to access records marked as `general' and `restricted';
    \item nominated representatives with \textit{General} permissions to access only records marked as `general';
    \item service providers in the \textit{General} list upload records marked as `general', and access only records marked as `general';
    \item service providers in the \textit{Restricted} list upload records marked as `restricted', and access records marked as both `general' and `restricted';
\end{itemize}

The correctness of a system not only depends on its verification (i.e. have we modelled the system right?), but also on its validation (i.e. have we modelled the right system?). As a validation step, we used two plug-ins of Rodin: ProB, to translate the model to B; and EventB2Java, to generate JML-annotated Java code from the model. ProB can be used as an animator of the specification. The animation in ProB is fully automatic. We animated our model in ProB finding no inconsistencies; adding a level of confidence on the model. We also used the EventB2Java tool to translate the last refinement of the model. The Java implementation of \mhr consists of 48 Java classes. There is a Java class that
implements JUnit test cases: 30 test cases manually written that conform to the My Health Record specification. The implementation passed all tests. Both the ProB animation and the description of the test cases (along with the implementation) can be found in \cite{mhr:model}.

\section{Related Work}
\label{sec:relatedWork}
There are as many different models of online health record systems around the world as there are healthcare systems. There are also many different approaches implemented by different countries that have been designed based on their own policy frameworks. In United Kingdom, the National Health Service implemented the Summary Care Record\cite{scr} (SCR). A SCR is an electronic consumer record that contains health record information. Once a consumer is registered with a GP, then the consumer's SCR is created automatically (consumers can opt-out). Records in the system cannot be fully deleted, although consumers can request for their information to be withheld from the system once opted out. SCRs can be accessed by service providers' stuff (as long as they are in charge of the consumer's healthcare and they access the information through a secure system). In Austria, ELGA\cite{web:elga, elga} (\textit{elektronische Gesundheitsakte}) is an information system that provides consumers and their healthcare entities with an access to their health records. Health records are created at various health facilities. ELGA networks the data and makes it available electronically. The system is available to everyone who is covered by the Austrian health care system, and it enables consumers to opt out. In Singapore, the National Electronic Health Record\cite{nehr} (NEHR) is a secure online system that collects summary of consumers health records across different service provider providers. This enables authorised healthcare professionals to have a holistic picture of your healthcare history. 

These systems have similarities to \mhr. The key feature of \mhr, and difference w.r.t. to other systems, is that consumers have access to and can fully \textit{control} what is stored on their systems. They can also decide which service providers and nominated representatives can access which records. A similar system is the one developed in Estonia. The Electronic Health Record system\cite{ehelathrecord} (e-Health Record) creates a common record every consumer can access online. Every Estonian is automatically included in the system from birth, and they are enabled to opt out at any time. In e-Health Record, consumers can restrict records from service providers, they can also grant access to other consumers (i.e. nominated representatives). Unlike \mhr, consumers cannot delete any records nor upload information (e.g. blood sugar values). Furthermore, no formal specification and verification has been performed on these systems.

Our work closely follows Role Based Access control\cite{Sandhu:96, Ferraiolo:07} (RBAC), which is a classic access model that uses the notion of users, roles and defines the privileges between those notions. Users (e.g. service providers) might have specific roles (e.g. \textit{General} access) that define how to access resources (e.g. general service providers can only access records marked as `general'). Formalisation of RBAC has been done in Z\cite{Power:09} and in B\cite{Huynh:12}. We inspired our work on these works, although, the work present in this paper is shaped by a specific domain following specific policy frameworks. SGAC\cite{sgac} is a healthcare access control model that manages the access to consumers'  records. The model has been formally verified with Alloy\cite{alloy} and ProB\cite{prob}. Our work is similar to SGAC as it is the specification and verification of a health record system. However, the target specification language is different: we used Event-B and the associated tools for verification and code generation. Furthermore, unlike SGAC, the specification present in this paper is not the specification of an ideal system rather a realistic one. We present a specification that closely follows the legislative framework that defines the system, without making any assumption. As a future work, we plan to incorporate some of the SGAC features, e.g. more fine grained access to records and conflict resolution, as these enhance \mhr. \textit{eXtensible Access Control Markup}\cite{xacml} (XACML) is an attribute-based access control standard. It evaluates access requests according to rules. A rule has a subject (e.g. a service provider), an action (e.g. access), a resource (e.g. a record), and a effect (\textit{permit} or \textit{deny}). The standard has been formalised in CSP\cite{Bryans:05} and VDM\cite{Bryans:10}. XACML can be used for \mhr, but the standard does not natively support inheritance of permissions (e.g. \textit{Restricted} permissions inherit \textit{General} permissions).  Azeem \textit{et al.}\cite{Azeem:14} present a specification of an e-Health system in Z. The system deals with the scheduling of GPs and patients. Their work has the same spirit as the one in this paper (even though the system is different). However, we present not just the specification, but also the verification of the system and its validation. 

\section{Conclusion}
\label{sec:conclusion}

Electronic personal health record systems play a key role in today's society. The benefits are clear: avoids healthcare fragmentation and duplication; supports accessibility and availability, allowing timely and instantly access to information; improves coordination and quality of service providers; improves readability and accuracy, reducing misinterpretation; among others. No wonder why it is one of the 14 Grand Challenges for Engineering in the 21st Century\cite{nae}.  These systems contain sensitive and confidential health information about consumers. As such, it is imperative to make sure that the right mechanisms for control and access are in place. Not just to ensure that health information is properly managed (which is of paramount importance), but also for users to gain confidence in the system: it would be useless if users did not trust it. This paper presents the formal specification and verification of the My Health Record System, the electronic personal health record system being used in Australia. We retrieved a set of control and access properties from the Australian legislation that defines the system. We formally modelled the system and its properties in Event-B and proved it correct. As a validation step, the model was animated by translating it to ProB. This gave us a layer of confidence in the model. We then generated JML-annotated Java code from the Event-B model and manually implemented JUnit test cases that conform to the system specification. An advantage of testing the specification is that it is done in terms of the application domain; on a much higher abstract level. Therefore, tests are relevant, as they exercise user needs and requirements, taking the system as a whole.  

Work in progress is aimed at (i) proposing a more fine-grained mechanism for accessing health information, so users are able to control not just how to disclose health information to service providers, but also to GPs within those service providers; (ii) opening an avenue for using formalism to verify legislation. Formal modelling and the need to make abstractions and refine them, help understand the system, while finding inconsistencies in it. As future work, we plan to formally model the Act that defines the regulations for physical health records and formally prove the consistency between the legislation and \mhr.



\bibliographystyle{IEEEtran}
\bibliography{IEEEabrv,bibl}

\end{document}